# The Unusual Nature of Confined Modes in a Chiral System

Benjamin W. Zingsem[1], Michael Farle[1,2], Robert L. Stamps[3], and Robert E. Camley[4]

**Nonreciprocity of spin wave propagation is a well-known consequence of antisymmetric exchange contributions possible in magnetic spin systems that lack inversion symmetry. In this case, it is possible for the energy of a state to depend on the sign of its momentum as $\omega(k) \neq \omega(-k)$. We discuss here the consequences of this nonreciprocity on counterpropagating travelling spin wave states. In a confined geometry we find states with well-defined nodes but with amplitudes that are modulated such that inversion symmetry of the mode profile is lost. This feature leads to the suggestion that additional features may become visible in, for example, ferromagnetic resonance studies of ferromagnetic micro-elements with DMI, allowing a quantification of the amplitude and direction of the DMI. Moreover, this interference between nonreciprocal modes forms the basis for a generalized concept of mode confinement.**

Affiliations

[1] Faculty of Physics and Center for Nanointegration (CENIDE), University Duisburg-Essen, Duisburg, 47057, Germany

[2] Center for Functionalized Magnetic Materials, Immanuel Kant Baltic Federal University, 236041 Kaliningrad, Russian Federation.

[3] SUPA, School of Physics and Astronomy, University of Glasgow, Glasgow G12 8QQ, UK

[4] Center for Magnetism and Magnetic Nanostructures, University of Colorado at Colorado Springs, Colorado Springs, Colorado 80918, USA

*Background*. The creation of chiral Dzyaloshinskiĭ Moriya [1] [2] interactions (DMI) in thin films, through interface effects, has intriguing and potentially useful consequences [3] [4] [5] [6] [7] [8] [9] [10] [11] [12] [13] [14] [15] [16] [17] [18] [19] [20] [21].

There are many important questions regarding the impact of defects [22], impurities, geometry [23] and the potentials for tuning the interaction through composition and materials choices [24] [25] [26] [27] [28] [29] [30] [31] [32] [33] [34] [35] [36] [37] [38] [39]. Examination of dynamics, including investigations of spin waves and ferromagnetic resonance, are proving to be very interesting and useful [40] [41] [42] [43] [44] [45] [46] [47] [48] [49] [50] [51] [52] [53] [54] [55] [56]. One important application is the measure of DMI from analysis of data from spin wave spectroscopies.

It is well known that gradient magnetisation terms can contribute additional chiral exchange energies into spin wave frequencies [50] [57]. These shift the usual ferromagnetic dispersion relation $\omega \sim k^2$ off of zero with a $\pm Dk$ contribution that, if measured, provides a value for the DMI constant D [58] [59] [60]. There is an interesting question that arises when thinking about finite geometries. What happens in a chiral system when spin waves are confined by edges? The idea is sketched in Fig 1b, where two counter propagating spin wave modes are created by reflection from two mirrors. The waves have the same energy, but have different wavelengths when travelling in opposite directions because of the nonreciprocity of the medium. Immediately there follows an important consequence for measurements. Without DMI, the largest signal one would measure in an FMR experiment is the spatially uniform resonance mode [61]. Standing modes are typically invisible, with free end boundary conditions, as there is no net magnetic moment for a driving microwave field to couple to.

Suppose that the waves are propagating along the x axis between two mirrors placed at x= 0 and x= L . Because of DMI, the dispersion relation for spin waves is not symmetrical about +$k_x$ and –$k_x$ propagation [58]. As a result, in real space the interference pattern, at any instant in time, produced by the counter propagating waves is not necessarily symmetrical about the midplane (x=25) as seen in the example in

Fig. 1c. **A striking consequence of this real space asymmetry for measurement is that resonance experiments will detect more modes when DMI is present than when DMI is not present.** The reason for this is now evident. The DMI reduces the symmetry of the system [62] allowing <u>all</u> the modes to couple to the uniform driving field found in the typical ferromagnetic resonance experiment. Some modes are more strongly driven than others, typically the modes which have frequencies close to what used to be the ferromagnetic resonance frequency.

Micromagnetic calculations [63], which will be described later in this paper in detail, confirm this for ferromagnetic resonance obtained from an elliptical Py dot with and without DMI. The structure with DMI reveals more strong absorption peaks than the same structure does without DMI due to the lack of symmetry in the eigenmode profiles.

The asymmetry in the mode profile follows directly from the superposition of two oppositely propagating waves. Because of the nonreciprocity, one does not get simple standing waves. In Fig. 1c we see that the profile of the wave shifts in time, although the nodes remain at constant positions. As can be easily appreciated, in the presence of DMI the instantaneous mode profile does not have a fixed symmetry, i.e. it is not even or odd about the midpoint.

The essential features of the confined modes in the DMI case can be understood in the following perturbation argument. Suppose that the DMI is weak, so that the difference in wavelengths for the counter propagating waves is small. In the absence of DMI, even and odd symmetry solutions for confined modes are expected with wavenumbers $k_n$ quantized as multiples $n$ of the inverse confinement size $L$, according to $k_n = \frac{n\pi}{L}$.

Consider, as an example, a solution with even symmetry along the *x* direction of the form

$$m(x,t) = A\exp(-i\omega t)[\exp(i\,k_n x) + \exp(-i\,k_n x)] \tag{1}$$

This wave will obey the ferromagnetic exchange dispersion $\omega = Jk_n^2$ where *J* is the exchange energy in suitable units. Now allow for small DMI, with the strength parametrized by *D*, there is non-reciprocity for propagation along *x*. Because the right and left wavectors are different, we describe the confined wave with the equation

$$m_D(x) = A\exp(-i\omega t)[\exp(i\,k_n^R x) + \exp(+i\,k_n^L x)] \tag{2}$$

where the right wavevector is $k_n^R = k_n + k'$ and the left one is $k_n^L = -(k_n - k')$. With DMI, the wave has frequency $\omega = J(k_n + k')^2 \pm D(k_n + k')$. For small *D*, $k' = \pm\frac{D}{2J}$, which is also small. With these substitutions, the confined wave now has the form,

$$m_D(x) = A\exp(i\,k'x - i\omega t)\cos(k_n x) \tag{3}$$

This has a familiar 'beat' frequency structure, with a long wavelength envelope containing a short wavelength oscillation. Another way to view this is as a generalized confinement in a moving reference frame, resulting in a system with fixed nodes and a wave moving at velocity $v_p = \frac{\omega}{k'}$. As seen from our one-dimensional model, described below, the time dependent oscillations of the mode amplitude with DMI present are more complex than in the *D=0* case, with a pronounced time-dependent asymmetry in the mode profile, as seen in Fig. 1c. In particular, one sees a constant node position but with the profile of the wave moving as a function of time. Thus the numerical results are consistent with the simple analytic picture obtained above.

*Resonance Calculations.* Additional insight into the asymmetries of the confined modes can be obtained using a one dimensional atomic toy model for a spin chain. Although this model is oversimplified, there are distinct advantages to exploring it:

1) The microscopic model does not require a transition to a continuum limit with its reliance on smooth and slow variations for the solutions;

2) There is no need to have additional assumptions about boundary conditions;

3) A comparison of the results from the atomic model with the micromagnetic model can validate the assumptions inherent in the micromagentic calculations.

In the toy model, the equations of motion for classical spins are integrated numerically for a finite chain of length *N*. The equations of motion are damped, and of the Landau-Liftshitz-Gilbert form:

$$\frac{d}{dt}\vec{m}_i = -\gamma \vec{m}_i \times \vec{h}_i + \alpha \vec{m}_i \times \frac{d}{dt}\vec{m}_i \tag{4}$$

where $\alpha$ is the Gilbert damping constant and $\gamma$ is the gyromagnetic ratio. The local field at site i is $\vec{h}_i = -\partial E/\partial \vec{m}_i$ where the energy E is given by

$$E = \sum_{\langle ij \rangle}[-J\vec{m}_i \cdot \vec{m}_j + D\hat{z} \cdot (\vec{m}_i \times \vec{m}_j)] + \sum_i \vec{m}_i \cdot B\hat{z} \tag{5}$$

where the sum is over nearest neighbours; the chain of spins lies along the *x* axis, and *B* is a magnetic field applied perpendicular to the chain. The numerical integration is performed using a Runge-Kutta method. We discuss now the dynamics simulated over $10^{10}$ time steps after an initial starting state of spins uniformly canted slightly away from equilibrium by an angle of 1°. We use the parameters of Jm = 20 kOe and B = 1 kOe.

A Fourier transform in time is performed and used to generate a frequency spectrum of the modes. The Fourier transform is calculated for a transverse component of a sum of amplitudes taken over each site, $s_x(t) = \sum_i m_x(i)$. Spatial profiles of the mode amplitudes are then extracted as the site-dependent Fourier amplitude associated with a particular frequency peak in the spectrum. Note that the only non-vanishing sum of components will appear for mode profiles that exhibit a net magnetisation in the oscillating transverse components. Without DMI only the uniform ferromagnetic resonance mode satisfies this condition [64].

An example of the spectra with and without DMI is shown in Fig. 2. The spectrum without DMI, shown in the top panel of Fig. 2 , has only one peak, located just below 3 GHz. Setting the DMI to a value of D/J = 0.1 shifts the frequency of the main intensity peak and creates three additional peaks as seen in the middle panel of Fig. 2. This value of DMI is not large enough to change the ground state appreciably and the appearance of new peaks is associated with an asymmetry in the profiles that gives rise to a nonzero time-dependent magnetisation in the oscillating modes.

The spatial Fourier amplitude profiles corresponding to the four peaks are shown in the right panel of Fig. 2. These amplitudes appear to approximate the standing wave mode profiles one expects for confined spin waves without DMI, but this is misleading. As shown earlier, the node position remains fixed, but the time-dependent profile moves with a constant velocity. The Fourier profiles measure an integration of the motion over time, so the nodes remain at their expected positions, but the Fourier profile completely misses the motion of the wave. There are other subtle asymmetries in the profile shapes, and dramatic effects on the relative intensities. In particular, the ferromagnetic resonance mode, which as discussed above normally corresponds to the mode with the largest fluctuating transverse magnetic moment. Instead with DMI we find that the largest intensity mode, at 2.58 GHz, is the one with a node near the centre of the spin chain.

We now present results for a more realistic situation with a micromagnetic simulation of confined modes. An elliptical disc geometry is used with the length = 200 nm, width = 100 nm and thickness = 10 nm. The program Mumax 3.9.1 [63] was used to numerically integrate Landau-Lifshitz-Gilbert equations with parameters appropriate to FeNe (Permalloy composition with $M_s = 7.96 \times 10^5$ A/m, $A_{ex} = 13 \times 10^{-12}$ J/m, $\alpha = 0.002$ ). The DMI parameter is $D = 8 \times 10^{-4}$ J/m$^2$. The magnetic field, $B = 400$ mT, is applied in the plane of the ellipse and along the long axis. The DMI is of the type that resembles an interfacial symmetry breaking, out of plane. This is assured by using mumax's *Dind*[1] parameter [62]. A simple calculation shows, that for this type of symmetry breaking, the direction of nonreciprocity is perpendicular to the magnetization direction and can be described according to [58].

Spectra and mode amplitudes are shown in Fig. 3, contrasting the dynamics found with and without DMI. The mode amplitudes at different times, along the center of the long axis, are plotted along with the bounding surfaces. These boundaries represent the extremal amplitudes reached by the wave during its oscillations. Movies illustrating the oscillations throughout a complete cycle can be found at [ *"Source for online movie material goes here"*].

An essential feature in the oscillations is that while the amplitude increases and decreases between the bounds, it also shifts to the left with time. One consequence of this is that there is a net magnetic moment oscillating at each resonance frequency that can in principle be coupled to an oscillating magnetic field. The strength of the coupling, and therefore the absorption intensities, will depend upon the non-trivial time dependence of the mode amplitude. It is for this reason that the relative intensities of the peaks shown in the upper part of Fig. 3 do not vary simply with the number of nodes when DMI is present.

---

[1] The same simulation has been performed using the *Dbulk* parameter to define the chiral energy density. Here the direction of nonreciprocity lies along the Magnetization axis. Results on this are shown in the supplementary section.

The dependence of confined mode intensities on the strength of the DMI for a ferromagnetic elliptical disk is shown in Fig. 4. The material parameters are the same as those used previously. The top panel shows the relative strengths of the confined modes as a function of DMI. It is interesting that as the DMI strength is increased, modes with different wavelength become evident in the response of the system. Sample spectra and mode intensity profiles are shown in the lower part of Fig. 4. The spectra for two different values of the DMI are shown in the lower left panel, and the corresponding mode profiles are shown on the right. The mode spectra and profiles were calculated using a magnetic pulse designed to excite all modes in the ellipse, and Fourier transforming the results in time in order to extract spectra. As a consequence of dipolar effects, edge localized modes can appear, as well as surface localized propagating spin waves. In the case studied here, surface wave effects are not present due to the thinness of the disk, but edge localized modes are observed. The profiles B0 and A0, shown in the lower right panel for zero DMI, correspond to the ferromagnetic resonance and edge modes, respectively.

For a moderate value 1.0 mJ/m$^2$ of DMI, modes appear to have shifted to lower frequencies, and new high order standing modes are visible as C1 and D1. The mode that has the highest intensity is the one closest to the original FMR frequency. The overall mode structure does not change, but the profile of lowest frequency 'edge' mode (A1) appears to more persistent. The most dramatic feature is the new high order mode C1 which without DMI is not visible in the spectrum. We conclude that DMI in the confined geometry will allow observation of this, and other standing modes that are normally not visible in resonance. The resulting frequencies and mode intensities will allow unambiguous values for DMI to be determined.

*Summary*. The picture usually considered for travelling waves confined by reflection is one where the energies are quantized according to a restricted set of wavelengths. The allowed wavelengths are determined by boundary conditions and specify the number and location of nodal lines and planes. We have re-examined this construction for the case of non-reciprocal propagation wherein waves at a given frequency have wavelengths that depend upon direction of travel. Non-reciprocity of this sort is, at least in the magnetic spin systems we consider, possible when the effective Hamiltonian does not possess inversion symmetry. We find that confinement in this more general case of non-reciprocal propagation produces a fixed nodal structure as found in the reciprocal case, but the wave amplitudes are modulated in time with a phase velocity determined by the difference in wavelengths between inequivalent propagation directions.

The features above provide a simple method for detecting the presence of, and measuring the magnitude and sign, of the Dyzaloshinskiĭ-Moriya Interactions in magnetic spin systems. With DMI, the mode patterns have asymmetric time-varying amplitudes, producing a time-varying net transverse magnetic moment, and resulting in new peaks in a ferromagnetic resonance experiment. Measurement of the frequencies of these modes therefore provides direct and unambiguous values for DMI. Our results suggest that these new modes can be observed even in the case of weak DMI, and therefore also should be visible for relatively thick ferromagnetic films for which the DMI arises at interfaces.

*Acknowledgements*. RLS was supported under EPSRC EP/M024423/1, and REC received support from SUPA. REC would like to acknowledge the hospitality of Glasgow University, where much of this work was done.

# References

[1]  I. E. Dzyaloshinski{\u\i}, "Thermodynamic Theory of "Weak" Ferromagnetism In Antiferromagnetic Substances," *Soviet Physics Jetp,* vol. 5, no. 6, pp. 1259-1272, December 1957.

[2]  T. Moriya, "Anisotropic Superexchange Interaction and Weak Ferromagnetism," *Phys. Rev.,* vol. 120, pp. 91-98, Oct 1960.

[3]  M. Bode, M. Heide, K. von Bergmann, P. Ferriani, S. Heinze, G. Bihlmayer, A. Kubetzka, O. Pietzsch, S. Blugel and R. Wiesendanger, "Chiral magnetic order at surfaces driven by inversion asymmetry," *Nature,* vol. 447, no. 7141, pp. 190-193, #may# 2007.

[4]  X. Yu, N. Kanazawa, W. Zhang, T. Nagai, T. Hara, K. Kimoto, Y. Matsui, Y. Onose and Y. Tokura, "Skyrmion flow near room temperature in an ultralow current density," *Nat Commun,* vol. 3, pp. 988--, #aug# 2012.

[5]  B. Van Waeyenberge, A. Puzic, H. Stoll, K. W. Chou, T. Tyliszczak, R. Hertel, M. Fahnle, H. Bruckl, K. Rott, G. Reiss, I. Neudecker, D. Weiss, C. H. Back and G. Schutz, "Magnetic vortex core reversal by excitation with short bursts of an alternating field," *Nature,* vol. 444, no. 7118, pp. 461-464, #nov# 2006.

[6]  A. Fert, V. Cros and J. Sampaio, "Skyrmions on the track," *Nat Nano,* vol. 8, no. 3, pp. 152-156, #mar# 2013.

[7]  Y. Tchoe and J. H. Han, "Skyrmion generation by current," *Phys. Rev. B,* vol. 85, p. 174416, May 2012.

[8] S. Mühlbauer, B. Binz, F. Jonietz, C. Pfleiderer, A. Rosch, A. Neubauer, R. Georgii and P. Böni, "Skyrmion Lattice in a Chiral Magnet," *Science,* vol. 323, no. 5916, pp. 915-919, 2009.

[9] J. Iwasaki, M. Mochizuki and N. Nagaosa, "Universal current-velocity relation of skyrmion motion in chiral magnets," *Nat Commun,* vol. 4, pp. 1463--, #feb# 2013.

[10] X. Z. Yu, Y. Onose, N. Kanazawa, J. H. Park, J. H. Han, Y. Matsui, N. Nagaosa and Y. Tokura, "Real-space observation of a two-dimensional skyrmion crystal," *Nature,* vol. 465, no. 7300, pp. 901-904, #jun# 2010.

[11] W. Jiang, P. Upadhyaya, W. Zhang, G. Yu, M. B. Jungfleisch, F. Y. Fradin, J. E. Pearson, Y. Tserkovnyak, K. L. Wang, O. Heinonen, S. G. E. te Velthuis and A. Hoffmann, "Blowing magnetic skyrmion bubbles," *Science,* vol. 349, no. 6245, pp. 283-286, 2015.

[12] T. Schneider, A. A. Serga, T. Neumann, B. Hillebrands and M. P. Kostylev, "Phase reciprocity of spin-wave excitation by a microstrip antenna," *Phys. Rev. B,* vol. 77, p. 214411, Jun 2008.

[13] K. Sekiguchi, K. Yamada, S. M. Seo, K. J. Lee, D. Chiba, K. Kobayashi and T. Ono, "Nonreciprocal emission of spin-wave packet in FeNi film," *Applied Physics Letters,* vol. 97, no. 2, 2010.

[14] K.-S. Ryu, L. Thomas, S.-H. Yang and S. Parkin, "Chiral spin torque at magnetic domain walls," *Nat Nano,* vol. 8, no. 7, pp. 527-533, #jul# 2013.

[15] R. Skomski, Z. Li, R. Zhang, R. D. Kirby, A. Enders, D. Schmidt, T. Hofmann, E. Schubert and D. J. Sellmyer, "Nanomagnetic skyrmions," *Journal of Applied Physics,* vol. 111, no. 7, 2012.

[16] S. Heinze, K. von Bergmann, M. Menzel, J. Brede, A. Kubetzka, R. Wiesendanger, G. Bihlmayer and S. Blugel, "Spontaneous atomic-scale magnetic skyrmion lattice in two dimensions," *Nat Phys,* vol. 7, no. 9, pp. 713-718, #sep# 2011.

[17] P. Ferriani, K. von Bergmann, E. Y. Vedmedenko, S. Heinze, M. Bode, M. Heide, G. Bihlmayer, S. Bl\"ugel and R. Wiesendanger, "Atomic-Scale Spin Spiral with a Unique Rotational Sense: Mn Monolayer on W(001)," *Phys. Rev. Lett.,* vol. 101, p. 027201, Jul 2008.

[18] A. Bogdanov and A. Hubert, "Thermodynamically stable magnetic vortex states in magnetic crystals," *Journal of Magnetism and Magnetic Materials,* vol. 138, no. 3, pp. 255-269, 1994.

[19] A. Fert and P. M. Levy, "Role of Anisotropic Exchange Interactions in Determining the Properties of Spin-Glasses," *Phys. Rev. Lett.,* vol. 44, pp. 1538-1541, Jun 1980.

[20] S. Rohart and A. Thiaville, "Skyrmion confinement in ultrathin film nanostructures in the presence of Dzyaloshinskii-Moriya interaction," *Phys. Rev. B,* vol. 88, p. 184422, Nov 2013.

[21] S. Pizzini, J. Vogel, S. Rohart, L. D. Buda-Prejbeanu, E. Ju\'e, O. Boulle, I. M. Miron, C. K. Safeer, S. Auffret, G. Gaudin and A. Thiaville, "Chirality-Induced Asymmetric Magnetic Nucleation in $\mathrm{Pt}/\mathrm{Co}/{\mathrm{AlO}}_{x}$ Ultrathin Microstructures," *Phys. Rev. Lett.,* vol. 113, p. 047203, Jul 2014.

[22] U. K. Roszler, A. N. Bogdanov and C. Pfleiderer, "Spontaneous skyrmion ground states in magnetic metals," *Nature,* vol. 442, no. 7104, pp. 797-801, #aug# 2006.

[23] J. Iwasaki, M. Mochizuki and N. Nagaosa, "Current-induced skyrmion dynamics in constricted geometries," *Nat Nano,* vol. 8, no. 10, pp. 742-747, #oct# 2013.

[24] L. Zhang, H. Han, M. Ge, H. Du, C. Jin, W. Wei, J. Fan, C. Zhang, L. Pi and Y. Zhang, "Critical phenomenon of the near room temperature skyrmion material FeGe," *Scientific Reports,* vol. 6, pp. 22397--, #feb# 2016.

[25] Y. Tokunaga, X. Z. Yu, J. S. White, H. M. Rønnow, D. Morikawa, Y. Taguchi and Y. Tokura, "A new class of chiral materials hosting magnetic skyrmions beyond room temperature," *Nature Communications,* vol. 6, pp. 7638--, #may# 2015.

[26] J. Torrejon, J. Kim, J. Sinha, S. Mitani, M. Hayashi, M. Yamanouchi and H. Ohno, "Interface control of the magnetic chirality in CoFeB/MgO heterostructures with heavy-metal underlayers," *Nat Commun,* vol. 5, pp. --, #aug# 2014.

[27] F. Jonietz, S. M{\"u}hlbauer, C. Pfleiderer, A. Neubauer, W. M{\"u}nzer, A. Bauer, T. Adams, R. Georgii, P. B{\"o}ni, R. A. Duine, K. Everschor, M. Garst and A. Rosch, "Spin Transfer Torques in MnSi at Ultralow Current Densities," *Science,* vol. 330, no. 6011, pp. 1648-1651, 2010.

[28] J. Cho, N.-H. Kim, S. Lee, J.-S. Kim, R. Lavrijsen, A. Solignac, Y. Yin, D.-S. Han, N. J. J. van Hoof, H. J. M. Swagten, B. Koopmans and C.-Y. You, "Thickness dependence of the interfacial Dzyaloshinskii-Moriya interaction in inversion symmetry broken systems," *Nat Commun,* vol. 6, pp. --, #jul# 2015.

[29] S. Li\'{e}bana-Vi{\~n}as, U. Wiedwald, A. Elsukova, J. Perl, B. Zingsem, A. S. Semisalova, V. Salgueiriño, M. Spasova and M. Farle, "Structure-Correlated

Exchange Anisotropy in Oxidized Co80Ni20 Nanorods," *Chemistry of Materials,* vol. 27, no. 11, pp. 4015-4022, 2015.

[30] G. Chen, J. Zhu, A. Quesada, J. Li, A. T. N'Diaye, Y. Huo, T. P. Ma, Y. Chen, H. Y. Kwon, C. Won, Z. Q. Qiu, A. K. Schmid and Y. Z. Wu, "Novel Chiral Magnetic Domain Wall Structure in $\mathrm{Fe}/\mathrm{Ni}/\mathrm{Cu}(001)$ Films," *Phys. Rev. Lett.,* vol. 110, p. 177204, Apr 2013.

[31] S. X. Huang and C. L. Chien, "Extended Skyrmion Phase in Epitaxial $\mathrm{FeGe}(111)$ Thin Films," *Phys. Rev. Lett.,* vol. 108, p. 267201, Jun 2012.

[32] X. Z. Yu, N. Kanazawa, Y. Onose, K. Kimoto, W. Z. Zhang, S. Ishiwata, Y. Matsui and Y. Tokura, "Near room-temperature formation of a skyrmion crystal in thin-films of the helimagnet FeGe," *Nat Mater,* vol. 10, no. 2, pp. 106-109, #feb# 2011.

[33] J.-H. Park, C. H. Kim, H.-W. Lee and J. H. Han, "Orbital chirality and Rashba interaction in magnetic bands," *Phys. Rev. B,* vol. 87, p. 041301, Jan 2013.

[34] K.-W. Kim, H.-W. Lee, K.-J. Lee and M. D. Stiles, "Chirality from Interfacial Spin-Orbit Coupling Effects in Magnetic Bilayers," *Phys. Rev. Lett.,* vol. 111, p. 216601, Nov 2013.

[35] W. M\"unzer, A. Neubauer, T. Adams, S. M\"uhlbauer, C. Franz, F. Jonietz, R. Georgii, P. B\"oni, B. Pedersen, M. Schmidt, A. Rosch and C. Pfleiderer, "Skyrmion lattice in the doped semiconductor ${\text{Fe}}_{1\ensuremath{-}x}{\text{Co}}_{x}\text{Si}$," *Phys. Rev. B,* vol. 81, p. 041203, Jan 2010.

[36] R. Salikhov, L. Reichel, B. Zingsem, F. M. R{\"o}mer, R.-M. Abrudan, J. Rusz, O. Eriksson, L. Schultz, S. F{\"a}hler, M. Farle and others, "Enhanced and Tunable Spin-Orbit Coupling in Tetragonally Strained Fe-Co-B Films," *arXiv preprint arXiv:1510.02624,* 2015.

[37] G. Giannopoulos, R. Salikhov, B. Zingsem, A. Markou, I. Panagiotopoulos, V. Psycharis, M. Farle and D. Niarchos, "Large magnetic anisotropy in strained Fe/Co multilayers on AuCu and the effect of carbon doping," *APL Mater.,* vol. 3, no. 4, pp. -, 2015.

[38] M. Heide, G. Bihlmayer and S. Bl\"ugel, "Dzyaloshinskii-Moriya interaction accounting for the orientation of magnetic domains in ultrathin films: Fe/W(110)," *Phys. Rev. B,* vol. 78, p. 140403, Oct 2008.

[39] R. Lo Conte, E. Martinez, A. Hrabec, A. Lamperti, T. Schulz, L. Nasi, L. Lazzarini, R. Mantovan, F. Maccherozzi, S. S. Dhesi, B. Ocker, C. H. Marrows, T. A. Moore and M. Kl\"aui, "Role of B diffusion in the interfacial Dzyaloshinskii-Moriya interaction in $\mathrm{Ta}/{\mathrm{Co}}_{20}\mathrm{F}{\mathrm{e}}_{60}{\mathrm{B}}_{20}/\mathrm{MgO}$ nanowires," *Phys. Rev. B,* vol. 91, p. 014433, Jan 2015.

[40] T. Schulz, R. Ritz, A. Bauer, M. Halder, M. Wagner, C. Franz, C. Pfleiderer, K. Everschor, M. Garst and A. Rosch, "Emergent electrodynamics of skyrmions in a chiral magnet," *Nat Phys,* vol. 8, no. 4, pp. 301-304, #apr# 2012.

[41] S. Woo, K. Litzius, B. Kruger, M.-Y. Im, L. Caretta, K. Richter, M. Mann, A. Krone, R. M. Reeve, M. Weigand, P. Agrawal, I. Lemesh, M.-A. Mawass, P. Fischer, M. Klaui and G. S. D. Beach, "Observation of room-temperature magnetic

skyrmions and their current-driven dynamics in ultrathin metallic ferromagnets," *Nat Mater,* vol. 15, no. 5, pp. 501-506, #may# 2016.

[42] S. Emori, U. Bauer, S.-M. Ahn, E. Martinez and G. S. D. Beach, "Current-driven dynamics of chiral ferromagnetic domain walls," *Nat Mater,* vol. 12, no. 7, pp. 611-616, #jul# 2013.

[43] A. Yamaguchi, T. Ono, S. Nasu, K. Miyake, K. Mibu and T. Shinjo, "Real-Space Observation of Current-Driven Domain Wall Motion in Submicron Magnetic Wires," *Phys. Rev. Lett.,* vol. 92, p. 077205, Feb 2004.

[44] A. Thiaville, S. Rohart, Ã. JuÃ©, V. Cros and A. Fert, "Dynamics of Dzyaloshinskii domain walls in ultrathin magnetic films," *EPL (Europhysics Letters),* vol. 100, no. 5, p. 57002, 2012.

[45] M. Uchida, Y. Onose, Y. Matsui and Y. Tokura, "Real-Space Observation of Helical Spin Order," *Science,* vol. 311, no. 5759, pp. 359-361, 2006.

[46] S. Emori, E. Martinez, K.-J. Lee, H.-W. Lee, U. Bauer, S.-M. Ahn, P. Agrawal, D. C. Bono and G. S. D. Beach, "Spin Hall torque magnetometry of Dzyaloshinskii domain walls," *Phys. Rev. B,* vol. 90, p. 184427, Nov 2014.

[47] N. Nagaosa and Y. Tokura, "Topological properties and dynamics of magnetic skyrmions," *Nat Nano,* vol. 8, no. 12, pp. 899-911, #dec# 2013.

[48] K. Zakeri, Y. Zhang, J. Prokop, T.-H. Chuang, N. Sakr, W. X. Tang and J. Kirschner, "Asymmetric Spin-Wave Dispersion on Fe(110): Direct Evidence of the Dzyaloshinskii-Moriya Interaction," *Phys. Rev. Lett.,* vol. 104, p. 137203, Mar 2010.

[49] SampaioJ., CrosV., RohartS., ThiavilleA. and FertA., "Nucleation, stability and current-induced motion of isolated magnetic skyrmions in nanostructures," *Nat Nano,* vol. 8, no. 11, pp. 839-844, #nov# 2013.

[50] V. E. Demidov, M. P. Kostylev, K. Rott, P. Krzysteczko, G. Reiss and S. O. Demokritov, "Excitation of microwaveguide modes by a stripe antenna," *Applied Physics Letters,* vol. 95, no. 11, 2009.

[51] S. D. Yi, S. Onoda, N. Nagaosa and J. H. Han, "Skyrmions and anomalous Hall effect in a Dzyaloshinskii-Moriya spiral magnet," *Phys. Rev. B,* vol. 80, p. 054416, Aug 2009.

[52] A. Terwey, R. Meckenstock, B. W. Zingsem, S. Masur, C. Derricks, F. M. Römer and M. Farle, "Magnetic anisotropy and relaxation of single Fe/FexOy core/shell-nanocubes: A ferromagnetic resonance investigation," *AIP Advances,* vol. 6, no. 5, 2016.

[53] L. Udvardi and L. Szunyogh, "Chiral Asymmetry of the Spin-Wave Spectra in Ultrathin Magnetic Films," *Phys. Rev. Lett.,* vol. 102, p. 207204, May 2009.

[54] A. T. Costa, R. B. Muniz, S. Lounis, A. B. Klautau and D. L. Mills, "Spin-orbit coupling and spin waves in ultrathin ferromagnets: The spin-wave Rashba effect," *Phys. Rev. B,* vol. 82, p. 014428, Jul 2010.

[55] S. Masur, B. Zingsem, T. Marzi, R. Meckenstock and M. Farle, "Characterization of the oleic acid/iron oxide nanoparticle interface by magnetic resonance," *Journal of Magnetism and Magnetic Materials ,* vol. 415, pp. 8-12, 2016.

[55] A. B. Butenko, A. A. Leonov, U. K. R\"o\ss{}ler and A. N. Bogdanov, "Stabilization of skyrmion textures by uniaxial distortions in noncentrosymmetric cubic helimagnets," *Phys. Rev. B,* vol. 82, p. 052403, Aug 2010.

[56] P. Khalili Amiri, B. Rejaei, M. Vroubel and Y. Zhuang, "Nonreciprocal spin wave spectroscopy of thin Ni–Fe stripes," *Applied Physics Letters,* vol. 91, no. 6, 2007.

[57] J.-H. Moon, S.-M. Seo, K.-J. Lee, K.-W. Kim, J. Ryu, H.-W. Lee, R. D. McMichael and M. D. Stiles, "Spin-wave propagation in the presence of interfacial Dzyaloshinskii-Moriya interaction," *Phys. Rev. B,* vol. 88, p. 184404, Nov 2013.

[58] A. A. Stashkevich, M. Belmeguenai, Y. Roussign\'e, S. M. Cherif, M. Kostylev, M. Gabor, D. Lacour, C. Tiusan and M. Hehn, "Experimental study of spin-wave dispersion in Py/Pt film structures in the presence of an interface Dzyaloshinskii-Moriya interaction," *Phys. Rev. B,* vol. 91, p. 214409, Jun 2015.

[59] D. Cortés-Ortuño and P. Landeros, "Influence of the Dzyaloshinskii–Moriya interaction on the spin-wave spectra of thin films," *Journal of Physics: Condensed Matter,* vol. 25, no. 15, p. 156001, 2013.

[60] M. Farle, "Ferromagnetic resonance of ultrathin metallic layers," *Reports on Progress in Physics,* vol. 61, no. 7, p. 755, 1998.

[61] A. N. Bogdanov and U. K. R\"o\ss{}ler, "Chiral Symmetry Breaking in Magnetic Thin Films and Multilayers," *Phys. Rev. Lett.,* vol. 87, p. 037203, Jun 2001.

[62] A. Vansteenkiste, J. Leliaert, M. Dvornik, M. Helsen, F. Garcia-Sanchez and B. Van Waeyenberge, "The design and verification of MuMax3," *AIP Advances,* vol. 4, no. 10, 2014.

[64] S. Vonsovski{\u\i}, Ferromagnetic resonance: the phenomenon of resonant absorption of a high-frequency magnetic field in ferromagnetic substances, Pergamon Press, 1966.

[65] S. Li\'{e}bana-Vi{\~n}as, R. Salikhov, C. Bran, E. M. Palmero, M. Vazquez, B. Arvan, X. Yao, P. Toson, J. Fidler, M. Spasova, U. Wiedwald and M. Farle, "Magnetic hardening of Fe 30 Co 70 nanowires," *Nanotechnology,* vol. 26, no. 41, p. 415704, 2015.

[66] T. Holstein and H. Primakoff, "Field Dependence of the Intrinsic Domain Magnetization of a Ferromagnet," *Phys. Rev.,* vol. 58, pp. 1098-1113, Dec 1940.

[67] R. Damon and J. Eshbach, "Magnetostatic modes of a ferromagnet slab," *Journal of Physics and Chemistry of Solids,* vol. 19, no. 3, pp. 308-320, 1961.

Figure 1

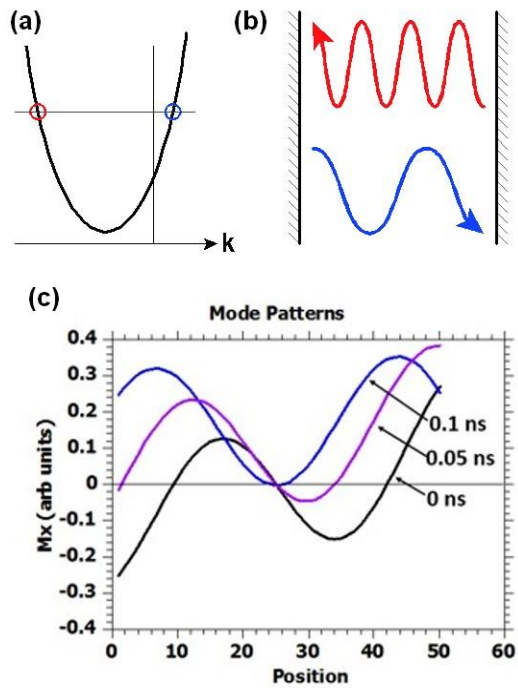

(a) Schematic of nonreciprocal dispersion relation when DMI is present. (b) The direction of propagation leads to different wavelengths. (c) Profiles of the spin excitations, calculated with the one-dimensional model, at different times for the mode at 2.58 GHz. We see a node position that is constant, but the profile of the mode slides to the left with time.

Figure 2

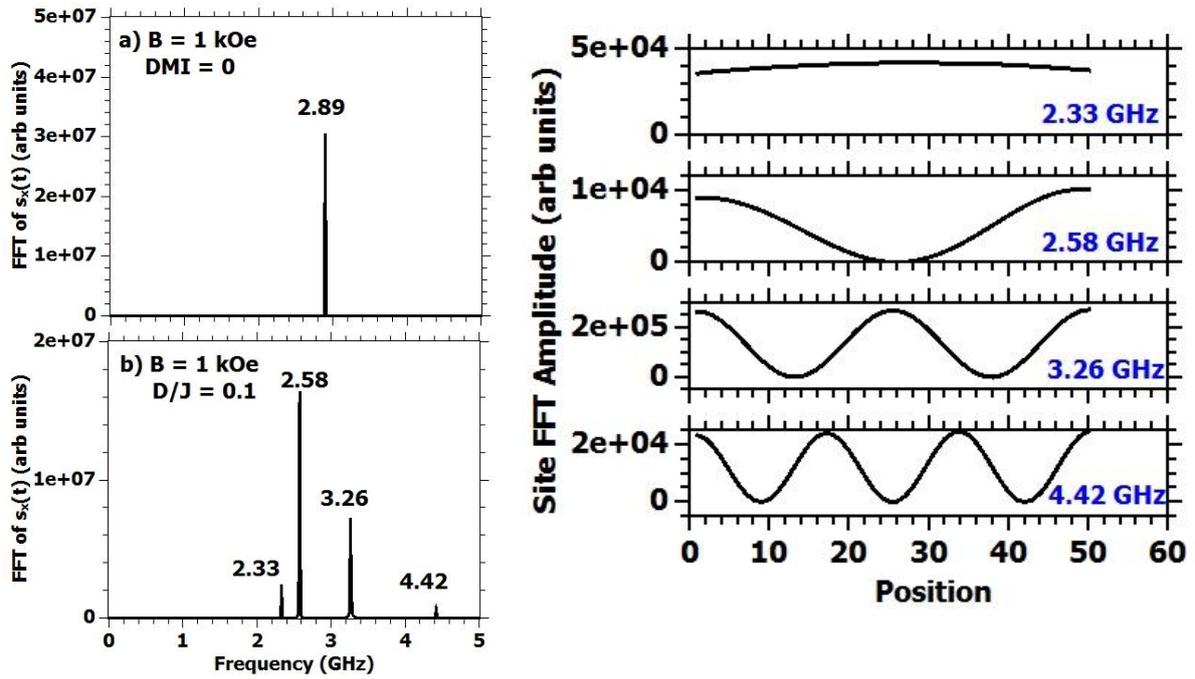

One dimensional spin chain simulation results for excitations caused by a small uniform canting from equilibrium. Comparison of frequencies calculated (a) without DMI (top) and (b) with DMI (bottom). The numbers indicate the mode frequencies in GHz. New modes are visible with DMI. (Right panel) Spatial amplitude profiles for the modes at different frequencies seen in the DMI calculations.

Figure 3

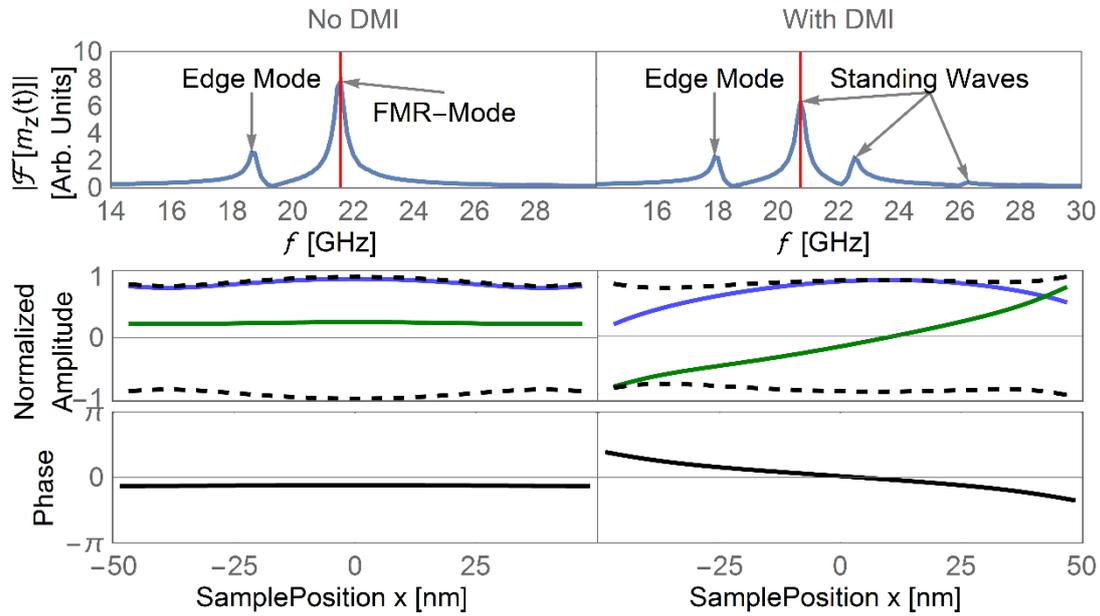

Results from micromagnetic simulations of spin waves confined in an elliptical disk. The FMR spectra without DMI (left) and with DMI (right) are shown at the top. They are represented by Fourier transformed of the time dependent z-component of the magnetization. The mode profiles, along the short axis of the ellipse, which correspond to the highest amplitudes in the spectra (each marked with a red vertical line) are shown in the lower section together with the phase angle. The outline of the mode profiles is indicated by dashed lines . Time snapshots of the wave are shown as green and blue curves. The green curve is plotted a quarter period later then the blue curve in the time evolution.

Figure 4

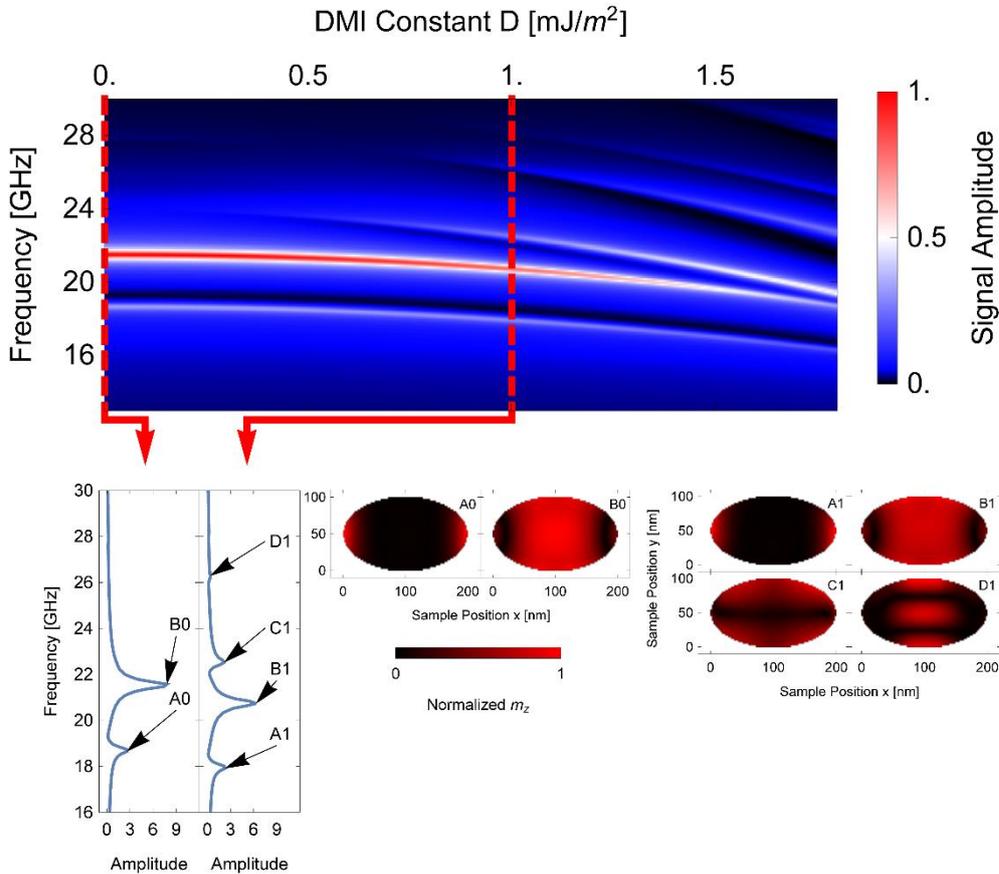

Micromagnetic calculations showing how mode intensities and profiles depend upon interfacial DMI for an elliptical ferromagnet. The upper plot summarizes the frequencies and intensities of modes as a function of the DMI strength. Shown in the lower left are mode spectra calculated by taking the Fourier transform in time after excitation by a shaped pulse. The corresponding mode intensity profiles are shown on the right and identified by labels A0, etc.

Supplementary Information

Supplementary Figure 1

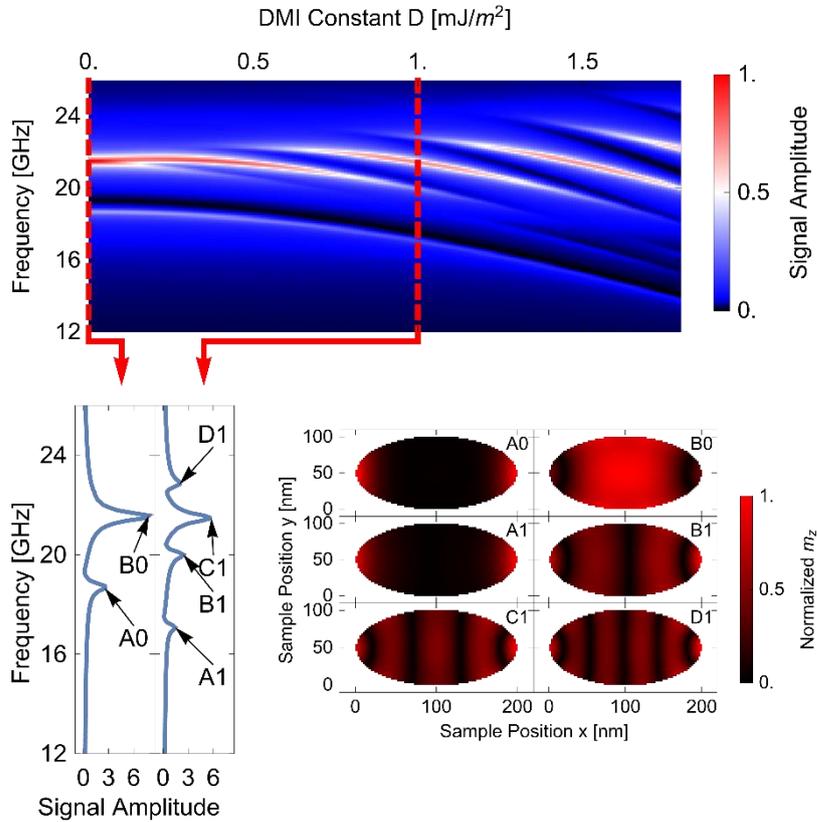

Evolution of the FMR spectrum under variation of the chiral energy density for Py parameters, with the magnetization and applied field parallel to the long axis of the ellipse. The direction of nonreciprocity is along the magnetization direction, confining more modes in a smaller frequency spectrum. The upper plot summarizes the frequencies and intensities of modes as a function of the DMI strength. Shown in the lower left are mode spectra calculated by taking the Fourier transform in time after excitation by a shaped pulse. The corresponding mode intensity profiles are shown on the right and identified by labels 1A, etc.

Supplementary Figure 2

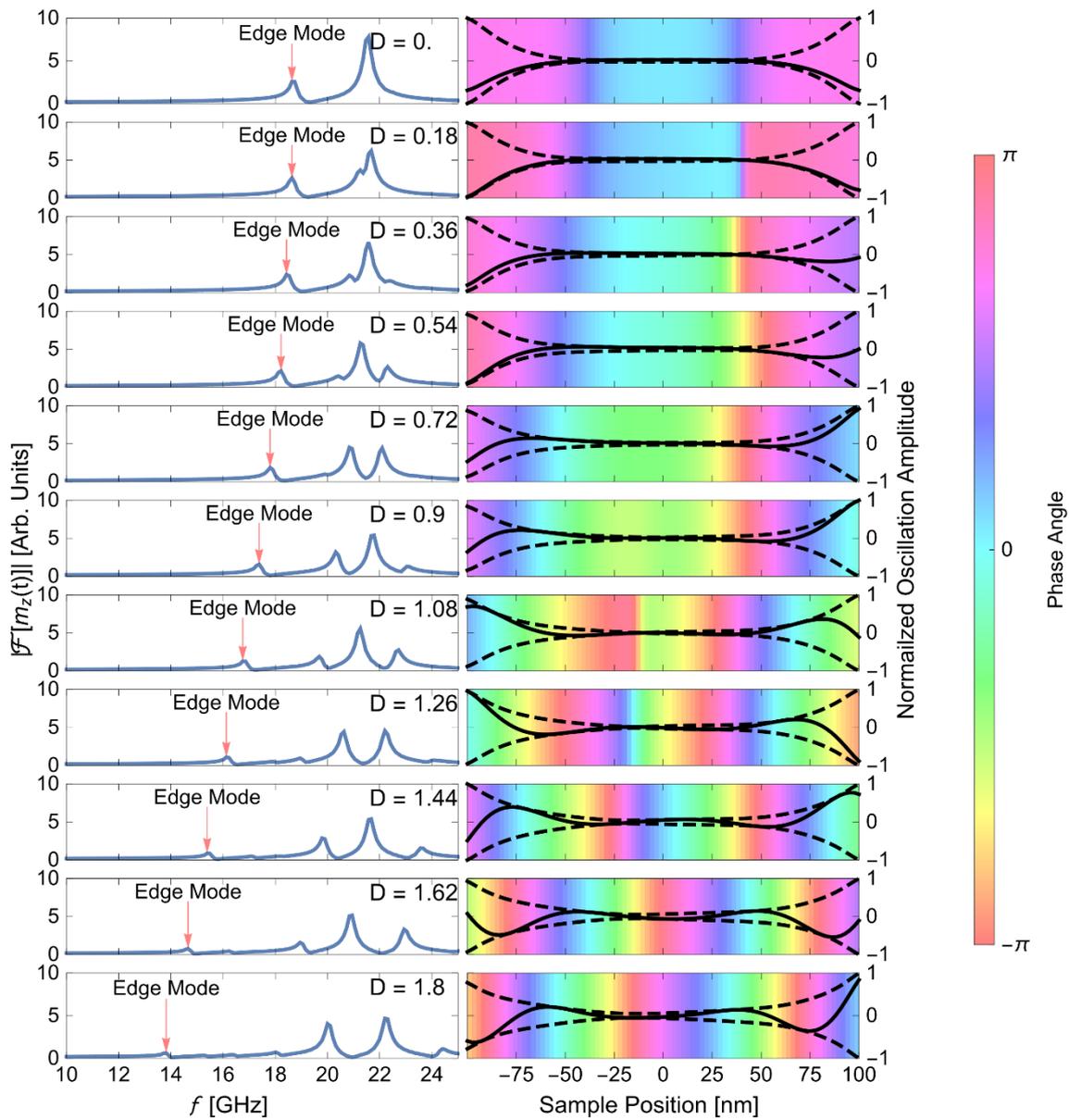

Evolution of the edge localized mode in Supplementary Figure 1 with increasing DMI. The left column shows absorption spectra at different DMI values D in mJ/m². The right column shows the mode profiles corresponding to the edge modes in the spectra. The black dashed lines indicate the wave amplitude, the black lines are time snapshots of the oscillation.